\def\isarxiv{1} 

\ifdefined\isarxiv
\documentclass[11pt]{article}

\usepackage[numbers]{natbib}

\else
\documentclass{article}
\usepackage{neurips_2022}
\fi

\usepackage{amsmath}
\usepackage{amsthm}
\usepackage{amssymb}
\usepackage{algorithm}
\usepackage{subfig}
\usepackage{algpseudocode}
\usepackage{graphicx}
\usepackage{grffile}
\usepackage{wrapfig,epsfig}
\usepackage{url}
\usepackage{xcolor}
\usepackage{epstopdf}

\usepackage{bbm}
\usepackage{dsfont}

\allowdisplaybreaks

\ifdefined\isarxiv

\usepackage{tikz}
\usepackage{hyperref}  
\hypersetup{colorlinks=true,citecolor=blue,linkcolor=blue} 
\usetikzlibrary{arrows}
\usepackage[margin=1in]{geometry}

\else

\usepackage{microtype}
\usepackage{hyperref}
\definecolor{mydarkblue}{rgb}{0,0.08,0.45}
\hypersetup{colorlinks=true, citecolor=mydarkblue,linkcolor=mydarkblue}

\fi

\newtheorem{theorem}{Theorem}[section]
\newtheorem{lemma}[theorem]{Lemma}
\newtheorem{definition}[theorem]{Definition}

\newtheorem{fact}[theorem]{Fact}

\newcommand{\wt}{\widetilde}

\newcommand{\R}{\mathbb{R}}

\DeclareMathOperator{\D}{\mathsf{D}}

\DeclareMathOperator{\poly}{poly}

\DeclareMathOperator{\nnz}{nnz}

\DeclareMathOperator{\diag}{diag}

\makeatletter
\newcommand*{\RN}[1]{\expandafter\@slowromancap\romannumeral #1@}
\makeatother

\usepackage{lineno}

\begin{document}

\ifdefined\isarxiv

\date{}

\title{Randomized and Deterministic Attention Sparsification Algorithms for Over-parameterized Feature Dimension\thanks{The authors thank Lichen Zhang for generous help and suggesting the major error control idea in the analysis.}}
\author{
Yichuan Deng\thanks{\texttt{ycdeng@cs.washington.edu}. The University of Washington.}
\and 
Sridhar Mahadevan\thanks{\texttt{smahadev@adobe.com}. Adobe Research.}
\and
Zhao Song\thanks{\texttt{zsong@adobe.com} Adobe Research.}
}

\else

\title{Intern Project} 
\maketitle 
\fi

\ifdefined\isarxiv
\begin{titlepage}
  \maketitle
  \begin{abstract}
Large language models (LLMs) have shown their power in different areas. Attention computation, as an important subroutine of LLMs, has also attracted interests in theory. Recently the static computation and dynamic maintenance of attention matrix has been studied by [Alman and Song 2023] and [Brand, Song and Zhou 2023] from both algorithmic perspective and hardness perspective. In this work, 
we consider the sparsification of the attention problem. We make one simplification which is the logit matrix is symmetric. 
Let $n$ denote the length of sentence, let $d$ denote the embedding dimension. 
Given a matrix $X \in \R^{n \times d}$, suppose $d \gg n$ and $\| X X^\top \|_{\infty} < r$ with $r \in (0,0.1)$, then we aim for finding $Y \in \R^{n \times m}$ (where $m\ll d$) such that
\begin{align*}
\| D(Y)^{-1} \exp( Y Y^\top ) - D(X)^{-1} \exp( X X^\top) \|_{\infty} \leq O(r)
\end{align*}
We provide two results for this problem.
\begin{itemize}
    \item Our first result is a randomized algorithm. It runs in $\widetilde{O}(\nnz(X) + n^{\omega} ) $ time, has $1-\delta$ succeed probability, and chooses $m = O(n \log(n/\delta))$. Here $\nnz(X)$ denotes the number of non-zero entries in $X$. We use $\omega$ to denote the exponent of matrix multiplication. Currently $\omega \approx 2.373$.
    \item Our second result is a deterministic algorithm. It  runs in $\wt{O}(\min\{\sum_{i\in[d]}\nnz(X_i)^2, dn^{\omega-1}\} + n^{\omega+1})$ time and chooses $m = O(n)$. Here $X_i$ denote the $i$-th column of matrix $X$.
\end{itemize}
Our main findings have the following implication for applied LLMs task: for any super large feature dimension, we can reduce it down to the size nearly linear in length of sentence.

  \end{abstract}
  \thispagestyle{empty}
\end{titlepage}

{
}
\newpage

\else

\begin{abstract}
Large language models (LLMs) have shown their power in different areas. Attention computation, as an important subroutine of LLMs, has also attracted interests in theory. Recently the static computation and dynamic maintenance of attention matrix has been studied by [Alman and Song 2023] and [Brand, Song and Zhou 2023] from both algorithmic perspective and hardness perspective. In this work, 
we consider the sparsification of the attention problem. We make one simplification which is the logit matrix is symmetric. 
Let $n$ denote the length of sentence, let $d$ denote the embedding dimension. 
Given a matrix $X \in \R^{n \times d}$, suppose $d \gg n$ and $\| X X^\top \|_{\infty} < r$ with $r \in (0,0.1)$, then we aim for finding $Y \in \R^{n \times m}$ (where $m\ll d$) such that
\begin{align*}
\| D(Y)^{-1} \exp( Y Y^\top ) - D(X)^{-1} \exp( X X^\top) \|_{\infty} \leq O(r)
\end{align*}
We provide two results for this problem.
\begin{itemize}
    \item Our first result is a randomized algorithm. It runs in $\widetilde{O}(\nnz(X) + n^{\omega} ) $ time, has $1-\delta$ succeed probability, and chooses $m = O(n \log(n/\delta))$. Here $\nnz(X)$ denotes the number of non-zero entries in $X$. We use $\omega$ to denote the exponent of matrix multiplication. Currently $\omega \approx 2.373$.
    \item Our second result is a deterministic algorithm. It  runs in $\wt{O}(\min\{\sum_{i\in[d]}\nnz(X_i)^2, dn^{\omega-1}\} + n^{\omega+1})$ time and chooses $m = O(n)$. Here $X_i$ denote the $i$-th column of matrix $X$.
\end{itemize}
Our main findings have the following implication for applied LLMs task: for any super large feature dimension, we can reduce it down to the size nearly linear in length of sentence.

\end{abstract}

\fi

\section{Introduction}

Attention mechanisms have become an essential tool in many natural language processing (NLP) applications, and large language models (LLMs). LLMs, for examples, Transformer \cite{vsp+17}, GPT-1 \cite{rns+18}, BERT \cite{dclt18}, GPT-2 \cite{rwc+19}, GPT-3 \cite{bmr+20}, PaLM \cite{cnd+22} and OPT \cite{zrg+22} have significantly advanced the state of the art in this field.  
These models rely on attention mechanisms to capture the dependencies and relations between different tokens in a sequence, making them ideal for tasks such as language modeling \cite{mms+19}, machine translation \cite{hwl21}, and sentiment analysis \cite{uas20}. A recent breakthrough, ChatGPT, which is a chatbot by OpenAI built with GPT-3, has shown the power of attention mechanism \cite{cha22}.  
Very recently, OpenAI released their technical report on GPT-4 \cite{o23}, in many real-life tasks it has significant better performance than previous versions of GPT.

In LLMs, attention mechanism is a fundamental component \cite{vsp+17, rns+18, dclt18, rwc+19, bmr+20}. Attention matrices are typically represented using an attention matrix $A \in \mathbb{R}^{n \times n}$, where $n$ is the length of the input sequence. Each element $A_{i,j}$ of the attention matrix represents the attention score or weight assigned to the $j$-th input element for the $i$-th output token. The attention scores are often computed using a softmax function applied to a query-key dot product matrix $QK^\top$, where $Q,K \in \mathbb{R}^{n \times d}$ are the query and key matrices, respectively, and $d$ is the dimensionality of the input embeddings. (Here we focus on the case when $d \gg n$.)
More precisely, the attention scores can be computed as $ A = \exp(QK^\top)$. 
The output of the attention mechanism is then computed as a weighted sum of the input sequence elements, with the attention scores serving as the weights. Formally, the attention matrix computation is defined as
\begin{definition}[Attention matrix computation]
    Let $Q, K \in \R^{n \times d}$ be two matrices. For a 
    matrix $V \in \R^{n \times d}$, the attention matrix is defined as
    \begin{align*}
        \mathrm{Att}(Q, K, V) := D^{-1}AV,
    \end{align*}
    where the matrices $A \in \R^{n \times n}$ and $D \in \R^{n \times n}$ are defined as
    \begin{align*}
        A := \exp(QK^\top) \text{~~and~~}
        D := \diag(A\mathbf{1}_n). 
    \end{align*}
\end{definition}

The attention computation is the major bottleneck for LLM models \cite{vsp+17, rns+18, dclt18, rwc+19, bmr+20, wlk+20, kkl20}. While attention mechanisms have shown remarkable performance on a wide range of NLP tasks, the computational cost of computing the attention matrix can be prohibitively expensive, especially for larger feature dimension. Naively computing the matrix will be expensive when the dimension is large. Since $A$ is in the shape of $n \times n$, naively it requires at least $\Omega(n^2)$ time to compute $D^{-1}AV$. However, under some specific precision guarantees, \emph{approximating} the attention computing suffices for most LLMs \cite{cgrs19, kkl20, wlk+20, dkod20, kvpf20, cdw+21, cdl+22}. 
A recent work \cite{as23} has studied when can we compute $\mathrm{Att}(Q, K, V)$ without \emph{explicitly} computing $A$ (with only access to matrices $Q, K, V$). They \cite{as23} provided two results, on the positive side, they show if the entries satisfy some property, then there is an algorithm that can compute the attention in $n^{1+o(1)}$ time. On the negative side, they show if the entries does not satsify that property, unless strong exponential time hypothsis is false, there is no truly subquadratic time to compute the attention approximately up to $1/\poly(n)$ additive error.  
In the work of \cite{bsz23}, they generalize the static attention computation to dynamic attention maintenance problem. They provide a conditionally\footnote{Their hardness result is based on a variation of OMV conjecture \cite{hkns15,lw17,ckl18} which is called Hinted MV conjecture \cite{bns19}.} optimal dynamic algorithm for attention maintenance if the Hinted MV conjecture is true.

Inspired by previous works and applications of attention, we study the following question,
\begin{center}
{\it
    Is it possible to compute the attention matrix faster for large feature dimension? 
}
\end{center}
To address this challenge, we seek for fast attention matrix computing algorithm. Here in this work, we focus on a specific form of attention matrix computation by assuming $X = Q = K$ (ignore the effect of $V$): 
\begin{definition}[Symmetric attention approximation]
    Let $X \in \R^{n \times d}$ be a matrix, the goal is to approximate
    \begin{align*}
        \D^{-1}(X)\exp(XX^\top),
    \end{align*}
    where $\D(X) := \diag(\exp(XX^\top)\mathbf{1}_n)$.
\end{definition}
For this specific setting, we provide an affirmative response to the above question in this work. 

\subsection{Our Results}
In this work, we present an algorithm for approximating symmetric attention matrices in a computationally efficient manner. The algorithm takes as input a large feature matrix $X$ with dimensions $n \times d$ with a constraint that $\| X X^\top \|_{\infty} \leq r$. It then outputs a compressed matrix $Y \in \R^{n \times m}$ with $m= O(n \log (n/\delta))$, where $\delta \in (0,1)$ is a failure probability that users can choose on their own in practice. The algorithm runs in nearly sparsity time of $X$ and outputs an approximation to the symmetric attention of $X$. 

Inspired by the approach of \cite{dsw22}, formally, we prove: 

\begin{theorem}[Randomized algorithm, informal version of Theorem~\ref{thm:main_randomized_formal}]\label{thm:main_randomized_informal}
    Let $r \in (0,0.1)$. 
    Given a matrix $X \in \R^{n \times d}$ with $d \gg n$ and $\| X X^\top \|_{\infty} \leq r$, there is a randomized algorithm that runs in time
    \begin{align*}
    \wt{O}(\nnz(X) + n^{\omega} ) 
    \end{align*}
    and outputs a matrix $Y \in \R^{n \times m}$ with $m= O(n \log (n/\delta))$ such that
    \begin{align*}
    \| \D(Y)^{-1} \exp(YY^\top) - \D^{-1}(X) \exp(XX^\top)\|_{\infty} \leq O(r)
    \end{align*}
    holds with probability $1-\delta$. 
    
    Here $\D(X) := \diag( \exp(XX^\top) {\bf 1}_n )$. We use $\omega$ to denote the exponent of matrix multiplication. Currently $\omega \approx 2.373$.
\end{theorem}

Following the trend of derandomized algorithm for linear programming by Brand's breakthrough result~\cite{b20}. In this work, we also consider how to design a deterministic sparsification algorithm for attention matrix. 
Inspired by another approach \cite{sxz22,z22} of deterministic sparsification method, we proposed another deterministic algorithm as follows. 

\begin{theorem}[Deterministic algorithm, informal version of Theorem~\ref{thm:main_deterministic_formal}]\label{thm:main_deterministic_informal}
    Let $r \in (0,0.1)$. 
    Given a matrix $X \in \R^{n \times d}$ with $d \gg n$ and $\| X X^\top \|_{\infty} \leq r$, there is a randomized algorithm that runs in time
    \begin{align*}
        \wt{O}(\min\{\sum_{i\in[d]}\nnz(X_i)^2, dn^{\omega-1}\} + n^{\omega+1})
    \end{align*}
    and outputs a matrix $Y \in \R^{n \times m}$ with $m= O(n)$ such that
    \begin{align*}
        \| \D(Y)^{-1} \exp(YY^\top) - \D^{-1}(X) \exp(XX^\top)\|_{\infty} \leq O(r). 
    \end{align*}
     
\end{theorem}
Unlike the result of \cite{b20} matching the linear programming algorithm by \cite{cls19}, here our deterministic algorithm is much slower than randomized algorithm. We are not sure if it's possible to design a deterministic algorithm which can match the time of our randomized algorithm. We leave this as an open problem for the future work.

\subsection{Related Work}

\paragraph{Theory of Attention Computation}
There is a series of work focused on Attention computation \cite{zhdk23,as23, bsz23,lsz23}. 
Studies have utilized locality sensitive hashing (LSH) techniques for attention approximation \cite{zhdk23}. With the observation that the kernel density estimation (KDE) problem can be used to simplify the denominator of the softmax function, the proposed the KDEformer \cite{zhdk23} which employed an efficient KDE solver so that it can approximate the attention in sub-quadratic time with provable spectral norm bounds. 
Furthermore, both static and dynamic versions of attention computation have been explored in recent work \cite{as23, bsz23}. The work of \cite{lsz23} studied the regularized hyperbolic regression problems such exponential function, cosh functions and sinh functions.

\paragraph{Input Sparsity Time Algorithm}

There has been a long history of the research in algorithms runs in input sparsity time \cite{cw13, nn13,rsw16,swz17,als+18,swz19, syyz22, dsw22,lsz23}. \cite{cw13} provides the first input sparsity time algorithm for linear regression and low-rank approximation. Given any input matrices $A$, the product of a sparse embedding matrix $S$ and $A$ can be computed in $\nnz(A)$ time. Later work \cite{nn13} gives Oblivious Sparse Norm Approximating Projections (OSNAPs), improving the  dimension in \cite{cw13}. Many works \cite{rsw16,swz17} are generalizing the Frobenius norm low-rank approximation to other norms. Recent work \cite{syyz22} proposed a faster algorithm for approximating the John Ellipsoid in input sparsity time. Another recent work \cite{dsw22} gave an algorithm such that, for any real matrix, it can solve the discrepancy minimization problem in the input sparsity time of the matrix.

\paragraph{Sketching and Sampling}

Sketching and sampling techniques are very powerful in numerical linear algebra. It has been applied to many fundamental tasks, Like linear programming (LP) \cite{cls19,jswz21,y20,dly21,gs22}, cutting plane method \cite{jlsw20}, empirical risk minimization \cite{lsz19,qszz23}, integral minimization \cite{jlsz23}, semi-definite programming \cite{syyz22_lichen}, matrix sensing \cite{dls23}, federated learning \cite{swyz22}, frank-wolfe method \cite{sxyz22}, matrix completion \cite{gsyz23}, dynamic sparsifier \cite{djs+22}, differential privacy \cite{emm+23}, clustering \cite{emz19,cmz22,emmz22}, tensor problems \cite{rsz22,syyz23}, training over-parameterized neural networks \cite{szz21,z22}.

\paragraph{Roadmap.}
We organize the paper as follows. In Section~\ref{sec:prel} we give the preliminary for our paper, including notations, algebraic facts and definition of sketching matrices. In Section~\ref{sec:ana} we provide the analysis of the correctness of our algorithms. In Section~\ref{sec:sparse} we provide our sparsification tools. In Section~\ref{sec:leverage_score} we give the analysis for the leverage score sampling tool we use.

\paragraph{Acknowledgements.}

The authors would like to thank Josh Alman, Jan van den Brand, Beidi Chen, Yeqi Gao, Zhihang Li, Junze Yin, Lichen Zhang, and Tianyi Zhou for very helpful discussions. 

\section{Preliminary}
\label{sec:prel}
In Section~\ref{sec:notation}, we provide the notations to use acress the paper. In Section~\ref{sec:basic_fact}, we state some well-known facts. In Section~\ref{sec:sketch_mat} we introduce some sketching matrices.

\subsection{Notations}
\label{sec:notation}
We use $[n]$ to denote the set $\{1,2,\cdots, n\}$ for any positive integer.  
For a matrix $A \in \R^{n \times n}$, we say $A$ is positive semidefinite (PSD) if $\forall x \in \R^n$, we have $x^\top A x \geq 0$. For any matrix $A \in \R^{n \times n}$, we use $\|A\|_F$ to denote its Frobenius norm, we use $\|A\|_1$ to denote its entry-wise $\ell_1$ norm, i.e., $\|A\|_F := (\sum_{i \in [n]} \sum_{j\in [n]} |a_{i,j}|^2 )^{1/2}$ and $\|A\|_1 := \sum_{i\in [n]} \sum_{j \in [n]} |a_{i,j}|$.  
We use $\| A \|_{\infty}$ to denote $\max_{i,j} | A _{i,j} |$.

For a matrix $A$, we use $\nnz(A)$ to denote the number of nonzero in a matrix $A$. 
For matrix $A$, we use $\| A \|$ to denote its operator/spectral norm, i.e., $\| A \| :=\max_x \|A x \|_2 / \|x \|_2$. Note that $\| A \|_2$ means nothing, such notation should never exist in the draft. 
For a PSD matrix $A$, let $U \Sigma U^\top$ denote its SVD. We use $A^{1/2}$ to denote $U \Sigma^{1/2} U^\top$.

The Taylor expansion of $\exp(x)$ is $\sum_{i = 0}^\infty \frac{x^i}{i!} = 1 + x + \frac{x^2}{2!} + \cdots$.

We use $\omega$ to denote the exponent of matrix multiplication, currently $\omega \approx 2.373$ \cite{w12,lg14,aw21}.  

\subsection{Basic Facts}
\label{sec:basic_fact}

\begin{fact}\label{fac:basic}
\begin{itemize}
    \item For any $x \in [-0.1,0.1]$,  $|1-\exp(x)| \leq 2 |x|$
\end{itemize}
\end{fact}
\begin{proof}
 
    By Taylor expansion, for $x \in [-0.1, 0.1]$, we have
    \begin{align*}
        |1 - \exp(x)| 
    = & ~   |\sum_{i = 1}^\infty\frac{x^i}{i!}| \\
    \le & ~ |x| + |\sum_{i = 2}^\infty\frac{x^i}{i!}| \\
    \le & ~ |x| + |\sum_{i = 2}^\infty(\frac{x^2}{2})^i| \\
    = & ~   |x| + |\frac{x}{2-2x}| \\
    \le & ~ 2|x|,
    \end{align*}
    where the 1st step follows from the Taylor expansion of $\exp(x)$, the 2nd step follows from the triangle inequality, the 3rd step follows from lower bounding $i!$ with $2^i$ since $i > 2$, the 4th step follows from summing up of the second term, and the last step follows from $|x| \le 0.1$. 
\end{proof}

\begin{fact}\label{fac:psd_diagonal_vs_off_diagonal}
Let $B \in \R^{n \times n}$ denote a psd matrix. Then we have 
\begin{align*}
B_{i,i} B_{j,j} \geq B_{i,j}^2, ~~~\forall i,j \in [n] \times [n]
\end{align*}
\end{fact}
\begin{proof}
We choose $x = a \cdot e_i + b \cdot e_j$.
Since for any $a,b \in \R$, then we have
\begin{align*}
0 \leq & ~ x^\top B x \\
= & ~ \begin{bmatrix} a & b \end{bmatrix} \begin{bmatrix} 
B_{i,i}  & B_{i,j}  \\
B_{j,i} & B_{j,j}
\end{bmatrix}
\begin{bmatrix} a \\ b \end{bmatrix}
\end{align*}
Since for any $a,b$ the above equation is true, thus 

\begin{align*}
\det (  \begin{bmatrix} 
B_{i,i}  & B_{i,j}  \\
B_{j,i} & B_{j,j}
\end{bmatrix} ) \geq 0
\end{align*}
which means $B_{i,i} B_{j,j} \geq B_{i,j}^2$.

\end{proof}

\begin{fact}\label{fac:A_B_psd_close}
For any $A \in \R^{n \times n}$ and $B \in \R^{n \times n}$ if the following condition holds
\begin{itemize}
    \item $(1-\epsilon) A \preceq B \preceq (1+\epsilon) A $
\end{itemize}
then we have
\begin{align*}
(1-\epsilon) A_{i,i} \preceq B_{i,i} \preceq (1+\epsilon) A_{i,i}, ~~~ \forall i \in [n].
\end{align*}
\end{fact}
\begin{proof}

For each $i \in [n]$, we use $e_i \in \R^n$ to denote the length-$n$ where $i$-th entry is $1$ and all the other entries are zeros.

We choose $x = e_i$. Then by
\begin{align*}
    (1-\epsilon) A \preceq B \preceq (1+\epsilon) A,
\end{align*}
we have
\begin{align*}
    (1-\epsilon) e_i^\top A e_i \preceq e_i^\top B e_i \preceq (1+\epsilon) e_i^\top A e_i,
\end{align*} 
which completes the proof.

\end{proof}

\subsection{Sketching Matrices}
\label{sec:sketch_mat}
Here we formally define several kinds of sketching matrices. 
\begin{definition}[Sparse Embedding Matrix I \cite{nn13}]
\label{def:sparse_trans_I}
    We say $R \in \R^{b \times n}$ is a sparse embedding matrix with parameter $s$ if each column has exactly $s$ non-zero elements being $\pm 1/\sqrt{s}$ uniformly at random, whose locations are picked uniformly at random without replacement (and independent across columns).
\end{definition}

\begin{definition}[Random Gaussian matrix or Gaussian transform, folklore]

\label{def:Gaussian_trans}
    Let $S = \sigma \cdot G \in \R^{s \times m}$ where $\sigma$ is a scalar, and each entry of $G \in \R^{s \times m}$ is chosen independently from the standard Gaussian distribution. For any matrix $A \in \R^{m \times n}$, $S A$ can be computed in $O(s \cdot \nnz(A))$ time.
\end{definition}

\begin{definition}[AMS \cite{ams96}]\label{def:AMS_trans}
    Let $h_1,h_2, \dots, h_b$ be $b$ random hash functions picking from a $4$-wise independent hash family $\mathcal{H} = \{h: [n] \rightarrow \{-\frac{1}{\sqrt{b}}, +\frac{1}{\sqrt{b}}\} \}$. Then $R \in \R^{b \times n}$ is a AMS sketch matrix if we set $R_{i,j} = h_i(j)$.
\end{definition}

\section{Analysis}
\label{sec:ana}
We consider a over-parameterized version attention matrix approximation problem. Suppose $d \gg n$. Here in this section, we provide analysis of the two kinds of the algorithms. In Section~\ref{sec:prop_psd} we prove some properties of PSD matrices to be used. In Section~\ref{sec:pert_exp_a_mat} we prove the perturbation of entry-wise exponentiating a matrix. In Section~\ref{sec:pert_diag_mat} we prove the perturbation of diagonal normalization matrix. In Section~\ref{sec:pert_att} we prove the perturbation of attention matrix. In Section~\ref{sec:rand_alg} and Section~\ref{sec:determin_alg} we provide our randomized algorithm and deterministic algorithm respectively.

\subsection{Property of a PSD Matrix}
\label{sec:prop_psd}

\begin{lemma}\label{lem:perturb_psd}
Let $A \in \R^{n \times n}$ and $B \in \R^{n \times n}$ denote two psd matrices.

If the following conditions hold
\begin{itemize}
    \item {\bf Condition 1.} $(1-\epsilon)B \preceq A \preceq (1+\epsilon) B$;
    \item {\bf Condition 2.} $-r \leq A_{i,j} \leq r$, for all $i,j \in [n] \times [n]$.
\end{itemize}
Then, we have
\begin{align*}
    B_{i,j} \in [ -(1+\epsilon)r, (1+\epsilon) r ].
\end{align*}
\end{lemma}
\begin{proof}

Using Fact~\ref{fac:A_B_psd_close} and {\bf Condition 1}, we know that
\begin{align}\label{eq:A_B_psd_close}
B_{i,i} \leq (1+\epsilon) A_{i,i}
\end{align}

We note that the matrix $B-(1-\epsilon)\cdot A$ is a PSD matrix, therefore,
\begin{align*}
    |B_{i,j}-(1-\epsilon)\cdot A_{i,j}|\leq & ~ \sqrt{(B_{i,i}-(1-\epsilon)\cdot A_{i,i})(B_{j,j}-(1-\epsilon)\cdot A_{j,j})} \\
    \leq & ~ 2\epsilon \sqrt{A_{i,i}A_{j,j}} \\
    \leq & ~ 2\epsilon r
\end{align*}
where the 1st step follows from Lemma~\ref{fac:psd_diagonal_vs_off_diagonal}, the 2nd step follows $B_{i,i} \leq (1 + \epsilon) A_{i,i}$ (see Eq.~\eqref{eq:A_B_psd_close}), and the 3rd step follows from the definition of $A_{i, j}$ from the lemma statement.

From Lemma {\bf Condition 2.},
\begin{align*}
    (1-\epsilon)\cdot A_{i,j} \in & ~ [- (1-\epsilon) \cdot r, (1-\epsilon)\cdot r],
\end{align*}

 Combining the above two equations, we obtain the following range on $B_{i,j}$:
\begin{align*}
    B_{i,j} \in & ~ [-(1+\epsilon)r , (1+\epsilon)r]
\end{align*}
We can do a symmetric argument using the fact that $(1+\epsilon)\cdot A-B$ is a PSD matrix:
\begin{align*}
    B_{i,j} \in & ~ [-(1+3\epsilon)r,(1+3\epsilon)r].
\end{align*}
The intersection of two ranges provides
\begin{align*}
    B_{i,j} \in & ~ [ - (1+\epsilon)r,(1+\epsilon)r]
\end{align*}

This completes the proof.
\end{proof}

\subsection{Perturbation of Entry-wise Exponentiating a Matrix} 
\label{sec:pert_exp_a_mat}
\begin{lemma}
\label{lem:perturb_exp}
    When the following conditions are true,
    \begin{itemize}
        \item {\bf Condition 1.} $r \in (0,0.1)$;
        \item {\bf Condition 2.} $A_{i,j} \in [-r, r]$;
        \item {\bf Condition 3.} $B_{i,j} \in [-(1+\epsilon)r, (1+\epsilon) r ]$.
    \end{itemize}
    Then, we have
    \begin{itemize}
    \item {\bf Part 1.} For all $i \in [n], j \in [n]$
    \begin{align*}
        | \exp(A_{i,j}) - \exp(B_{i,j}) | \leq \exp(A_{i,j}) \cdot 6 r.
    \end{align*}
    \item {\bf Part 2.} For all $i \in [n], j \in [n]$
    \begin{align*}
        | \exp( A_{i,j} ) - \exp( B_{i,j} ) | \leq \exp( B_{i,j} ) \cdot 6 r
    \end{align*}
    \end{itemize}
\end{lemma}
\begin{proof}
From {\bf Condition 2.} and {\bf Condition 3.}, we have
\begin{align}\label{eq:bound_A_i_j_minus_B_i_j}
    | A_{i,j} - B_{i,j} | \leq (2+\epsilon) r \leq 3 r.
\end{align}
where the 2nd step follows from $\epsilon \in (0,1)$.

{\bf Proof of Part 1.}
    We have
    \begin{align*}
    | \exp(A_{i,j}) - \exp(B_{i,j}) |
    = & ~ \exp(A_{i,j}) \cdot | 1 - \exp(B_{i,j} - A_{i,j}) | \\
    \leq & ~ \exp(A_{i,j}) \cdot \max\{ |1 - \exp(3r)| , |1-\exp(-3r)| \} \\
    \leq & ~ \exp(A_{i,j}) \cdot 6r. 
    \end{align*}
    where the 2nd step follows from Eq.~\eqref{eq:bound_A_i_j_minus_B_i_j}, the last step follows from {\bf Condition 1} in the Lemma statement and Fact~\ref{fac:basic}.

{\bf Proof of Part 2.}

We have
    \begin{align*}
    | \exp(A_{i,j}) - \exp(B_{i,j}) |
    = & ~ \exp(B_{i,j}) \cdot | 1 - \exp(A_{i,j} - B_{i,j}) | \\
    \leq & ~ \exp(B_{i,j}) \cdot \max\{ |1 - \exp(3r)| , |1-\exp(-3r)| \} \\
    \leq & ~ \exp(B_{i,j}) \cdot 6r. 
    \end{align*}
    where the 2nd step follows from Eq.~\eqref{eq:bound_A_i_j_minus_B_i_j}, and the last step follows from {\bf Condition 1} in the Lemma statement and Fact~\ref{fac:basic}.
\end{proof}

\subsection{Perturbation of Diagonal Normalization Matrix}
\label{sec:pert_diag_mat}

\begin{lemma}[Perturbation of diagonal normalization matrix]\label{lem:perturb_D}
If the following condition holds
\begin{itemize}
    \item {\bf Condition 1.} $| \exp(A_{i,j}) - \exp(B_{i,j}) | \leq \exp(A_{i,j}) \cdot 6 r$ for all $i \in [n], j \in [n]$
    \item {\bf Condition 2.} $| \exp(A_{i,j}) - \exp(B_{i,j}) | \leq \exp(B_{i,j}) \cdot 6 r$ for all $i \in [n], j \in [n]$
\end{itemize}
Then, we have,
\begin{itemize}
\item {\bf Part 1.}
For all $i \in [n]$
\begin{align*}
    | ( \exp( A_{i,*} ) - \exp(B_{i,*}) ) \cdot {\bf 1}_n | \leq \exp(A_{i,*}) {\bf 1}_n \cdot 6 r
\end{align*}
\item {\bf Part 2.} 
For all $i \in [n]$
\begin{align*}
    | ( \exp( A_{i,*} ) - \exp(B_{i,*}) ) \cdot {\bf 1}_n | \leq \exp(B_{i,*}) {\bf 1}_n \cdot 6 r
\end{align*}
\end{itemize}
\end{lemma}
\begin{proof}

{\bf Proof of Part 1.}
We can show that
\begin{align*}
| ( \exp( A_{i,*} ) - \exp(B_{i,*}) ) \cdot {\bf 1}_n | 
= & ~ | \sum_{j=1}^n ( \exp(A_{i,j}) - \exp(B_{i,j}) ) | \\
\leq & ~ \sum_{j=1}^n | \exp( A_{i,j} ) - \exp( B_{i,j} ) | \\
\leq & ~ \sum_{j=1}^n \exp(A_{i,j}) \cdot 6 r \\
= & ~ \exp(A_{i,*}) {\bf 1}_n \cdot 6 r
\end{align*} 
where the 1st step follows from simple algebra, the 2nd step follows from triangle inequality, the 3rd step follows from {\bf Condition 1.} in Lemma statement, and the last step follows from simple algebra.

{\bf Proof of Part 2.}
We can show that
\begin{align*}
| ( \exp( A_{i,*} ) - \exp(B_{i,*}) ) \cdot {\bf 1}_n | 
= & ~ | \sum_{j=1}^n ( \exp(A_{i,j}) - \exp(B_{i,j}) ) | \\
\leq & ~ \sum_{j=1}^n | \exp( A_{i,j} ) - \exp( B_{i,j} ) | \\
\leq & ~ \sum_{j=1}^n \exp(B_{i,j}) \cdot 6 r \\
= & ~ \exp(B_{i,*}) {\bf 1}_n \cdot 6 r
\end{align*} 
where the 1st step follows from simple algebra, the 2nd step follows from triangle inequality, the 3rd step follows from {\bf Condition 2.} in Lemma statement, and the last step follows from simple algebra.

\end{proof}

\subsection{Perturbation of Attention Matrix}
\label{sec:pert_att}
Here we state the perturbation lemma for attention matrix as follows. 
\begin{lemma}
\label{lem:perturb_attention}
Let $c_1 > 0$ and $c_2 >0$ denote two fixed constants.
    If the following conditions hold
    \begin{itemize}
        \item {\bf Condition 1.} For all $i \in [n]$,
        \begin{align*}
            | \D(X)_{i,i} - \D(Y)_{i,i} | \leq c_1 \cdot r \cdot \min \{ \D(X)_{i,i}, \D(Y)_{i,i} \}
        \end{align*}
    \item {\bf Condition 2.} For all $i \in [n]$, $j \in [n]$
        \begin{align*}
            | \exp((XX^\top)_{i,j}) - \exp((YY^\top)_{i,j}) | \leq c_2 \cdot r \cdot \min\{ \exp((XX^\top)_{i,j}), \exp((YY^\top)_{i,j}) \}
        \end{align*}
    \end{itemize}
    Then, we have
    \begin{align*}
        \|\D^{-1}(X)\exp(XX^\top) - \D^{-1}(Y)\exp(YY^\top)\|_\infty \le (c_1+c_2) \cdot r.
    \end{align*}
\end{lemma}

\begin{proof}
    We first decompose the difference into
    \begin{align*}
        & ~ \|\D^{-1}(X)\exp(XX^\top) - \D^{-1}(Y)\exp(YY^\top)\|_\infty \\
    \le & ~ \|\D^{-1}(X)\exp(YY^\top) - \D^{-1}(Y)\exp(YY^\top)\|_\infty + \|\D^{-1}(X)\exp(XX^\top) - \D^{-1}(X)\exp(YY^\top)\|_\infty \\
    = & ~ B_1 + B_2
    \end{align*}
where the last step follows from
\begin{align*}
    B_1 := \|\D^{-1}(X)\exp(YY^\top) - \D^{-1}(Y)\exp(YY^\top)\|_\infty,
\end{align*}
and
\begin{align*}
    B_2 := \|\D^{-1}(X)\exp(XX^\top) - \D^{-1}(X)\exp(YY^\top)\|_\infty.
\end{align*}

    We divide the proof into the two terms. 
    \paragraph{The first term.}
    For each of the index pair $(i, j) \in [n] \times [n]$, we have
    \begin{align*}
    B_1 = & ~ |(\D^{-1}(X)\exp(YY^\top) - \D^{-1}(Y)\exp(YY^\top))_{i, j}| \\
    =  & ~ | ( \D^{-1}(X)_{i,i} - \D^{-1}(Y)_{i,i})\exp(YY^\top)_{i,j}| \\
    =  & ~ | \frac{\D(X)_{i,i}-\D(Y)_{i,i}}{\D(X)_{i,i}\D(Y)_{i,i}}\exp(YY^\top)_{i,j}| \\
    \leq & ~ | \frac{\D(X)_{i,i}-\D(Y)_{i,i}}{\D(X)_{i,i}\D(Y)_{i,i}} | \cdot |   \exp(YY^\top)_{i,j} | \\
     \leq & ~ | \frac{c_{1} r \D(X)_{i,i}}{\D(X)_{i,i}\D(Y)_{i,i}} | \cdot |   \exp(YY^\top)_{i,j} | \\
    = & ~ c_1 r \cdot |\D(Y)^{-1}_{i,i} | \cdot |  \exp(YY^\top)_{i,j}| \\
    = & ~ c_1 r \cdot |\D(Y)^{-1}_{i,i} \exp(YY^\top)_{i,j}| \\
    = & ~ c_1 r \cdot (\D(Y)^{-1} \exp(YY^\top) )_{i,j} \\
    \leq & ~ c_1 r
    \end{align*}
    where the first step follows from definition, the second steps follow from simple algebra, the third step follows from triangle inequality, the forth step follows from  {\bf Condition 1} in the lemma statement, the fifth step follows from simple algebra, the sixth step follows from all the entries are positive, and the last step follows from definition of $\D$.

    \paragraph{The second term.}
    For each of the index pair $(i, j) \in [n] \times [n]$, we have
    \begin{align*}
     B_2 
     = & ~ |(\D^{-1}(X)\exp(XX^\top) - \D^{-1}(X)\exp(YY^\top))_{i, j}| \\
    = & ~ | \D^{-1}(X)_{i,i}\cdot(\exp(XX^\top)_{i,j} - \exp(YY^\top)_{i, j})| \\
    \leq & ~ \D^{-1}(X)_{i,i} \cdot | (\exp(XX^\top)_{i,j} - \exp(YY^\top)_{i, j})| \\
    \leq & ~ \D^{-1}(X)_{i,i} \cdot c_2 \cdot r \cdot (\exp(XX^\top)_{i,j}  \\
    \le & ~ c_2 r \cdot (\D^{-1}(X)\exp(XX^\top) )_{i,j} \\
    \leq & ~ c_2 r,
    \end{align*}
    where the first step follows from simple algebra, the second step follows from triangle inequality, the third step follows from  {\bf Condition 2} in the lemma statement, the forth step follows from simple algebra, and the last step follows from the definition of $\D$. 

    Finally, we have
     \begin{align*}
        \|\D^{-1}(X)\exp(XX^\top) - \D^{-1}(Y)\exp(YY^\top)\|_\infty 
        \leq & ~ B_1 + B_2 \\
        \leq & ~ ( c_1 +c_2 ) r.
    \end{align*}
    
    Thus we complete the proof. 
\end{proof}

\subsection{Randomized Symmetric Attention Approximation Algorithm}
\label{sec:rand_alg}

\begin{algorithm}\caption{Our Randomized Algorithm}
\label{alg:randomized}
\begin{algorithmic}[1]
\Procedure{Randomized}{$X \in \R^{n \times d}$,  $\delta$} \Comment{Theorem~\ref{thm:main_randomized_formal}}
    \State $\epsilon_\sigma \gets \Theta(1)$
    \State $\wt{\sigma} \gets \textsc{LeverageScore}(X, \epsilon_\sigma, \delta/2)$ \Comment{$\wt{\sigma} \in \R^n$, Algorithm~\ref{alg:fast_leverage_score}}
    \State $\epsilon \gets \Theta(1)$
    \State $m \gets \Theta( \epsilon^{-2} n \log(n/\delta) )$
    \State Sample a subset of $m$ rows of $X$ with proper re-weighting with respect to the leverage scores $\wt{\sigma}$
    \State Generate diagonal sampling matrix $\wt{D} \in \R^{n \times n}$ with respect to the above selection of rows
    \State \Return $Y = \wt{D}X$ \Comment{$Y \in \R^{n \times d}$}
\EndProcedure 
\end{algorithmic}
\end{algorithm}

\begin{theorem}[Randomized algorithm, formal version of Theorem~\ref{thm:main_randomized_informal}]
\label{thm:main_randomized_formal}
    Let $r \in (0,0.1)$. 
    Given a matrix $X \in \R^{n \times d}$ with $d \gg n$ and $\| X X^\top \|_{\infty} \leq r$, there is a randomized algorithm that runs in time
    \begin{align*}
    (\nnz(X) + n^{\omega} ) \cdot \poly(\log(n/\delta))
    \end{align*}
    and outputs a matrix $Y \in \R^{n \times m}$ with $m= O(n \log (n/\delta))$ such that
    \begin{align*}
    \| \D(Y)^{-1} \exp(YY^\top) - \D^{-1}(X) \exp(XX^\top)\|_{\infty} \leq O(r)
    \end{align*}
    holds with probability $1-\delta$. 
    
    Here $\D(X) = \diag( \exp(XX^\top) {\bf 1}_n )$.
\end{theorem}

\begin{proof}

{\bf Proof of Running Time.}
The running time follows from Lemma~\ref{lem:subsample}.

{\bf Proof of Correctness.}
    The correctness follows from Lemma~\ref{lem:rand_samp}, Lemma~\ref{lem:perturb_psd}, Lemma~\ref{lem:perturb_exp}, Lemma~\ref{lem:perturb_D}, and Lemma~\ref{lem:perturb_attention}. 
\end{proof}

\subsection{Deterministic Symmetric Attention Approximation Algorithm}
\label{sec:determin_alg}

\begin{theorem}[Deterministic algorithm, formal version of Theorem~\ref{thm:main_deterministic_informal}]
\label{thm:main_deterministic_formal}
    Let $r \in (0,0.1)$. 
    Given a matrix $X \in \R^{n \times d}$ with $d \gg n$ and $\| X X^\top \|_{\infty} \leq r$, there is a deterministic algorithm that runs in time
    \begin{align*}
        \wt{O}(\min\{\sum_{i\in[d]}\nnz(X_i)^2, dn^{\omega-1}\} + n^{\omega+1})
    \end{align*}
    and outputs a matrix $Y \in \R^{n \times m}$ with $m= O(n)$ such that
    \begin{align*}
        \| \D(Y)^{-1} \exp(YY^\top) - \D^{-1}(X) \exp(XX^\top)\|_{\infty} \leq O(r). 
    \end{align*}
    
    Here $\D(X) = \diag( \exp(XX^\top) {\bf 1}_n )$.
\end{theorem}

\begin{proof}

{\bf Proof of Running Time.}
The running time follows from Theorem~\ref{thm:determin_samp}.

{\bf Proof of Correctness.}
    The correctness follows from Theorem~\ref{thm:determin_samp}, Lemma~\ref{lem:perturb_psd}, Lemma~\ref{lem:perturb_exp}, Lemma~\ref{lem:perturb_D}, and Lemma~\ref{lem:perturb_attention}. 
\end{proof}

\section{Sparsification}
\label{sec:sparse}
In this section, we provide two kinds of sparsification tools. In Section~\ref{sec:rand_sparse} we provide our randomized sparsification algorithm and in Section~\ref{sec:determin_sparse} we provide the deterministic sparsification algorithm. 
\subsection{Randomized Sparsification Algorithm}
\label{sec:rand_sparse}
We have the following previous result.

\begin{lemma}
\label{lem:rand_samp}
    Given a $X \in \R^{n \times d}$. Let $m = \Theta( \epsilon^{-2} n \log(n/\delta) )$. Let $D \in \R^{m \times n}$ be the sampling matrix output by Algorithm~\ref{alg:randomized}. Let $Y = DX$ then we have
    \begin{align*}
        (1-\epsilon) XX^\top \preceq Y Y^\top \preceq (1+\epsilon ) X X^\top 
    \end{align*}
    holds with probability $1-\delta$.
\end{lemma}
\begin{proof}
    The proof follows from Lemma~\ref{lem:subsample}. 
\end{proof}

\subsection{Deterministic Sparsification Algorithm}
\label{sec:determin_sparse}
We have the following previous result. 
\begin{theorem}[Theorem~3.1 in \cite{sxz22}]
\label{thm:determin_samp}
    Let $X = \{x_1, \dots, x_d\} \subseteq \R^n$ be a set of vectors such that $\sum_{i=1}^d x_ix_i^\top = I_n$. Then there exists a deterministic algorithm such that, it can find a set of weights $\{w_i\}_{i=1}^d \subseteq \R$ satisfying
    \begin{align*}
        (1-\epsilon)\preceq\sum_{i=1}^d w_ix_ix_i^\top \preceq (1+\epsilon)I
        ~~\text{and}~~
        |\{i \in [d]:w_i\not=0\}| = \Theta(\epsilon^{-2}n).
    \end{align*}
    This algorithm can run in time
    \begin{align*}
        \wt{O}(\min\{\sum_{i\in[d]}\nnz(x_i)^2, dn^{\omega-1}\} + \epsilon^{-2}n^{\omega+1}). 
    \end{align*}
\end{theorem}

\section{Leverage Score Sampling}
\label{sec:leverage_score}
Here in this section, we provide a sampling algorithm based on \emph{leverage score}. In Section~\ref{sec:leverage_compute} we provide our fast leverage score computation algorithm. In Section~\ref{sec:time_correct_rand_sparse} we provide the running time and correctness proof for our randomized sparsification algorithm. In Section~\ref{sec:chernoff_sampling} we provide our sampling process with its analysis.

\subsection{Fast Leverage Score Computation}
\label{sec:leverage_compute}
In this section, we provide an algorithm which can fast approximate the leverage scores of an input matrix. 

\begin{algorithm}[!ht]\caption{Leverage Score Computation}\label{alg:fast_leverage_score}
\begin{algorithmic}[1]
\Procedure{LeverageScore}{$X \in \R^{n \times d}, \epsilon_{\sigma}=\Theta(1) ,\delta_{\sigma}$}\Comment{Lemma~\ref{lem:fast_leverage_score}}
    \State \Comment{This algorithm is designed for generating an $O(1)$ approximation to the leverage score}
    \State $s_1 \gets \wt{O}( \epsilon_{\sigma}^{-2} d)$
    \State Generate $S_1 \in \R^{s_1 \times n}$ as a sparse embedding matrix \Comment{Definition~\ref{def:sparse_trans_I}}
    \State $M \gets (S_1 \cdot X)$ \Comment{This step takes time of $\wt{O}( \epsilon_{\sigma}^{-1} \nnz(X)) $.}
    \State Let $R \in \R^{d \times d}$ denote the R of QR factorization of $M$ \label{line:fast_leverage_QR_M}
    \State $s_2 \gets \Theta(\log(n/\delta_{\sigma}))$
    \State Generate $S_2 \in \R^{d \times s_2 }$ as a JL matrix \Comment{Either Definition~\ref{def:Gaussian_trans} or Definition~\ref{def:AMS_trans}}
    \State Compute $N \gets R^{-1} S_2$ \label{line:fast_leverage_N_compute} \Comment{$N \in \R^{d \times s_2}$}
    \For{$j = 1 \to n$}\label{line:fast_leverage_sigma_loop} \Comment{This step takes time of $\wt{O}(\nnz(X))$.}
        \State $\wt{\sigma}_j \gets \| (e_j^\top X) N \|_2^2$
    \EndFor 
    \State \Return $\wt{\sigma}$ \Comment{$\wt{\sigma}_j = \Theta(\sigma_j), \forall j \in [n]$}
\EndProcedure 
\end{algorithmic} 
\end{algorithm}

Previous work \cite{dsw22} studied harder version of the leverage score sampling algorithm, which implies the following result directly.
\begin{lemma}[Leverage score, \cite{dsw22} ]\label{lem:fast_leverage_score}
    Given a matrix $X \in \R^{n \times d}$ there is an algorithm (procedure \textsc{LeverageScore} Algorithm~\ref{alg:fast_leverage_score}) that runs in $\wt{O}(\epsilon_{\sigma}^{-2}(\nnz(X) + n^{\omega}))$ time and output a vector $\wt{\sigma} \in \R^n$, such that, $\wt{\sigma}$ is an approximation of the leverage score of matrix $X$, i.e.,
    \begin{align*}
        \wt{\sigma} \in (1 \pm \epsilon_{\sigma}) \cdot \sigma(X),
    \end{align*}
    with probability at least $1 - \delta_{\sigma}$. The $\wt{O}$ hides the $\log(d/\delta_{\sigma})$ factors.
  
\end{lemma}

\subsection{Correctness and Running time}
\label{sec:time_correct_rand_sparse}

Here we present the correctness running time of our randomized sparsification algorithm based on the leverage score approximation algorithm.

\begin{lemma}\label{lem:subsample}
    There is an algorithm (Algorithm~\ref{alg:randomized}) takes $X \in \R^{n \times d}$ as input and outputs a diagonal sampling matrix $\wt{D}$ with $\| \wt{D} \|_0 = O( \epsilon^{-2} d \log(d/\delta) )$ and runs in time
    \begin{align*}
       \wt{O}( \nnz(X) + d^{\omega}),
    \end{align*}
    and it holds that
    \begin{align*}
        \Pr[ X^\top \wt{D}^\top \wt{D} X \in (1\pm \epsilon) \cdot X^\top X ] \geq 1-\delta. 
    \end{align*}
    Here $\wt{O}$ hides the $\log(d/\delta)$ factors. Here we use $\omega$ to denote the exponent of matrix multiplication. For current FMM algorithm, we have $\omega \approx 2.373$.
\end{lemma}
\begin{proof}
    We first approximately compute the leverage score, i.e., gives an $O(1)$-approximation to all leverage scores via Lemma~\ref{lem:fast_leverage_score}. Then run we samples a number of rows according to the leverage scores, using Lemma~\ref{lem:tilde_H}, we can show the correctness.
\end{proof}

\subsection{Sampling a Batch of Rank-\texorpdfstring{$1$}{} Terms}
\label{sec:chernoff_sampling}

In this section, we present the sampling process for sparsification of symmetric matrix in the shape of $A^\top A$. 
First, we define the following sampling process,
\begin{definition}[Sampling process for symmetric matrix]
\label{def:sample_process}
    Let $H=A^\top  A$. Let $p_j\geq {\beta\cdot\sigma_j(A)}/{n}$ for some $\beta \ge 1$. We sample $T$ rows of matrix $A$ with respect to the probability $p_j$ with replacement. We use $j_t$ to represent the index of the row that is sampled during the $t$-th trial. And the sampled matrix is defined as
    \begin{align*}
        \wt{H} := & ~ \frac{1}{T} \sum_{t=1}^T \frac{1}{p_{j_t}} a_{j_t}a_{j_t}^\top,
    \end{align*}
    where $T$ denotes the number of the trials. 
\end{definition}

For the above sampling process, previous work provided the following guarantee. 
\begin{lemma}[Sampling via Matrix Chernoff, see Lemma~5.2 in \cite{dsw22} as an example]
\label{lem:tilde_H}
    Let $\epsilon_0 \in (0, 1)$ denote the precision parameter, and $\delta_0 \in (0, 0.1)$ denote the failure probability. For the matrix $\wt{H}$ sampled as Definition~\ref{def:sample_process}, it holds with probability at least $1 - \delta_0$ that
    \begin{align*}
        (1-\epsilon_0)\cdot H \preceq \wt{H} \preceq (1+\epsilon_0)\cdot H.
    \end{align*}
    And the number of trials is
    \begin{align*} 
        T = \Theta( \epsilon_0^{-2}n\log(n/\delta_0)).
    \end{align*}
\end{lemma}

\ifdefined\isarxiv
\bibliographystyle{alpha}
\bibliography{ref}
\else
\bibliography{ref}
\bibliographystyle{alpha}

\fi

\newpage
\onecolumn
\appendix




\end{document}